# Suppression of antiferromagnetic spin fluctuations in the collapsed tetragonal phase of CaFe$_2$As$_2$

D.K. Pratt[1,2], Y. Zhao[3], S.A.J. Kimber[4], A. Hiess[5], D.N. Argyriou[4], C. Broholm[3], A. Kreyssig[1,2], S. Nandi[1,2], S.L. Bud'ko[1,2], N. Ni[1,2], P.C. Canfield[1,2], R.J. McQueeney[1,2] and A.I. Goldman[1,2]

[1]Ames Laboratory, US DOE, Iowa State University, Ames, IA 50011, USA

[2]Department of Physics and Astronomy, Iowa State University, Ames, IA 50011, USA

[3]Department of Physics & Astronomy, Johns Hopkins University, Baltimore, MD 21218 USA

[4]Helmholtz-Zentrum Berlin für Materialien und Energie, Glienicker Str. 100, 14109 Berlin, Germany

[5]Institut Laue-Langevin, 38042 Grenoble Cedex, France

## Abstract

Inelastic neutron scattering measurements of CaFe$_2$As$_2$ under applied hydrostatic pressure show that the antiferromagnetic spin fluctuations observed in the ambient pressure, paramagnetic, tetragonal (T) phase are strongly suppressed, if not absent, in the collapsed tetragonal (cT) phase. These results are consistent with a quenched Fe moment in the cT phase and the strong decrease in resistivity observed upon crossing the boundary from the T to cT phase. The suppression or absence of static antiferromagnetic order and dynamic spin fluctuations in the non-superconducting cT phase supports the notion of a coupling between spin fluctuations and superconductivity in the iron arsenides.

Although there is, as yet, no consensus regarding the pairing mechanism that gives rise to superconductivity in the iron arsenides, a great deal of activity has focused on the role of spin fluctuations and the proximity of superconductivity to structural and magnetic instabilities. In particular, recent inelastic neutron scattering measurements on polycrystalline $Ba_{0.6}K_{0.4}Fe_2As_2$ [1] as well as single crystals of $Ba(Fe_{0.92}Co_{0.08})_2As_2$ [2] and $Ba(Fe_{0.95}Ni_{0.05})_2As_2$ [3] have found evidence of a resonant spin excitation at energies above a spin gap that opens in the superconducting state. These resonances are found at energies corresponding to approximately $5k_BT_c$, similar to those observed for the high $T_c$ cuprates.[4] Furthermore, single crystal measurements [2,3] have shown that this resonance is found at the same wave vector, ($\frac{1}{2}$ $\frac{1}{2}$ $L$) for $L$ odd, as the static magnetic order in the undoped $AE$Fe$_2$As$_2$ ($AE$ = Ca, Sr, Ba) parent compounds, [5,6,7,8] again highly reminiscent of the cuprates. The interplay between spin dynamics and superconductivity has also been reported for intermetallic compounds such as $UPd_2Al_3$ [9], $CeCoIn_5$ [10], and $CeCu_2Si_2$ [11]. All of these results suggest that spin fluctuations are coupled to charge carriers, and may play an important role for superconductivity, in the iron arsenide compounds.

Superconductivity in the parent $AE$Fe$_2$As$_2$ compounds under applied pressure has been reported by several groups using liquid-media pressure cells. [12-18] For $CaFe_2As_2$, however, recent measurements [19] under hydrostatic pressure conditions (He pressure cell) have revealed that superconductivity, if present at all, occurs in a very narrow range of pressure close to the reported orthorhombic-to-collapsed tetragonal (O-cT) phase transition (see Fig. 1). [20,21] This work concluded that a non-hydrostatic pressure component may cause the formation of domains and domain walls with varying properties, including superconductivity.[19] Therefore, while the conditions necessary for superconductivity in $BaFe_2As_2$ and $SrFe_2As_2$ remain an open question, it appears that, under hydrostatic conditions, the cT phase in $CaFe_2As_2$ does not support superconductivity. There is also evidence that the cT phase in $CaFe_2As_2$, in contrast to the T and Orthorhombic (O) phases, is non-magnetic. [20,21] Elastic neutron scattering measurements under hydrostatic pressure (He pressure cell) , for example, have shown that the ordered moment of approximately 0.8 $\mu_B$ found for the antiferromagnetic O structure below $p = 0.35$ GPa disappears abruptly at the O-cT transition.[21] Furthermore, there is a striking increase in the conductivity of $CaFe_2As_2$ upon traversing the tetragonal-to-collapsed tetragonal (T-cT) phase boundary [12,13,19] even though band structure calculations show that the density of states at the Fermi energy is reduced by nearly 60% in the cT phase.[17] One possible explanation for this discrepancy is the elimination of spin scattering contributions to the resistivity on transforming from the paramagnetic T phase to a non-magnetic cT phase.

Here, we report on inelastic neutron scattering from spin fluctuations in $CaFe_2As_2$ under hydrostatic pressure. In the tetragonal phase we observe antiferromagnetically correlated spin fluctuations with a characteristic wave vector, $\mathbf{Q} = (½ ½ L)$ for $L$ odd, which are strongly suppressed or completely absent in the cT phase. This observation provides new evidence for the suppression of the Fe moment and, therefore, the loss of antiferromagnetic spin disorder scattering in the cT phase of $CaFe_2As_2$. Furthermore, the observed suppression of antiferromagnetic spin fluctuations in the non-superconducting cT phase supports the notion of strong coupling between spin fluctuations and superconductivity in the iron arsenide compounds.

For the inelastic measurements approximately 300 Sn-flux grown single crystals of $CaFe_2As_2$ [22] with a total mass of 1.5 grams were co-aligned on both sides of a set of aluminum plates (see Fig. 1(a)) such that their common [$H$ $H$ 0] and [0 0 $L$] axes (using the tetragonal unit cell notation) were coincident with the scattering plane. One of the challenges in this measurement lies in the nature of the T-cT transition in $CaFe_2As_2$. As described in references 20 and 21, there is a striking change in the unit cell dimensions upon transformation to the cT phase; the $c$ lattice parameter decreases by 10% while the $a$ lattice parameter increases by 2%. Our experience has shown that because $CaFe_2As_2$ is quite soft, constrained samples exhibit an extended coexistence between the cT and O or T phases. This coexistence is also observed when non-hydrostatic pressure is applied. [21] Therefore, only a light coating of fluorine-based (Fomblin) oil was used to maintain contact between the Al plates and the crystals. However, we have also found that this method can lead to changes in the alignment of crystals or the detachment and loss of some crystals, at the cT transition. The Al plates holding the samples were stacked and placed in the well of an Al-alloy He-gas pressure cell to ensure hydrostatic pressure conditions. The cell was connected to a pressure intensifier through a high pressure capillary that allowed continuous monitoring and adjustment of the pressure. Using this system, the pressure could be varied at fixed temperatures for $T > 50$ K (above the He solidification line), or the temperature could be changed at nearly constant pressures. A helium reservoir allowed the pressure to remain relatively constant as temperature was changed.

The inelastic neutron scattering measurements were performed on the IN8 triple-axis spectrometer at the Institut Laue-Langevin at a fixed final energy of 14.7 meV employing a double focusing Si(111) monochromator, a double focusing PG(002) analyzer and open collimation. A graphite filter was used after the sample to reduce harmonic contamination of the beam. The energy resolution in this configuration, measured using a vanadium standard, was 1.1 meV full-width-at-half-maximum (FWHM). The sample mosaic with respect to both the [$H\,H\,0$] and [$0\,0\,L$] directions was initially 1.6 deg FWHM at ambient pressure and increased slightly to 2 deg at $p = 0.5$ GPa. The pressure cell was loaded into an ILL Orange-type cryostat allowing temperature control from 300 K down to approximately 2 K.

Figures 1(b) and 1(c) describe the regions of reciprocal space and the *p-T* phase diagram investigated in these measurements. The inelastic scattering was studied primarily around the (½ ½ 1) antiferromagnetic wave vector (using the indices appropriate to the tetragonal unit cell). Constant energy **q**-scans were measured at several energies along both the [$H\,H\,0$] and [$0\,0\,L$] directions in order to characterize the spin fluctuations at selected temperatures and pressures. The sequence of measurements is depicted in Fig. 1(c). Since we have already established that crossing the phase boundaries into the cT phase from either the O or T phase can significantly increase the mosaic of the sample [17], care was taken to first measure the inelastic scattering at point 1 (ambient pressure in the tetragonal phase) and point 2 ($p = 0.5$ GPa but still in the tetragonal phase) before lowering temperature and crossing over into the cT phase itself (point 3). The scans at *p*

= 0.5 GPa were repeated using the same pressure cell containing only the aluminum slides coated with the fluorine-based oil and the He pressure medium to determine the background contribution to the observed scattering. Measurements of the spin wave dispersion in the O phase at ambient pressure ($T$ = 140 K) were also done prior to pressurizing the system for comparison with previous measurements by McQueeney *et al* [23] and are consistent with those results. Furthermore, no evidence of static magnetic order at the antiferromagnetic Bragg point in the T or cT phases was observed, consistent with previous results. [20, 21, 23]

Fig. 2 displays the raw data from constant energy **q**-scans through the (½ ½ 1) reciprocal lattice position along the [$H H$ 0] direction at ħω = 3 and 7 meV energy loss. This is the same wave vector where correlated spin fluctuations have been observed in the Co- and Ni-doped BaFe$_2$As$_2$ compounds. [2,3] The open squares in this Figure denote the measured, empty can, background data. Figures 2(a) and (b) clearly show the presence of correlated spin fluctuations centered on the antiferromagnetic (½ ½ 1) wave vector for both 3 and 7 meV. As temperature was lowered, at a constant pressure of 0.5 GPa, the sample transformed from the higher temperature T phase to the lower temperature cT phase at approximately 120 K. The absence of a clear signal above the empty can background in Figs. 2(c) and (d) indicates that the antiferromagnetic spin fluctuations in the cT phase (point 3 in Fig. 1(c)) are strongly suppressed, if not completely absent. Unfortunately, in the process of transforming from the T to the cT phase, there was an unavoidable increase in the sample mosaic from approximately 2 deg FWHM at 180 K and 0.5 GPa, to nearly 5 deg FWHM at 100 K and 0.5 GPa. Furthermore, the integrated intensity of the nuclear peaks measured in rocking scans decreased by approximately 40%, indicating that at least a portion of the sample was either dislodged or tilted well beyond the measured mosaic (this was verified by visual inspection of the sample after completion of the experiment). Therefore, the temperature was increased back to 180 K, to investigate the impact of the broadened mosaic and decreased sample volume on our ability to observe the spin fluctuations. Crossing the cT-T phase line again increased the sample mosaic to nearly 6.5 deg but did not affect the integrated intensities of the nuclear peaks. Figs. 2(e) and (f) demonstrate that although the magnetic scattering is broadened and decreased relative to the first measurement at 180 K and 0.5 GPa, it remains clearly in evidence, re-affirming that the correlated spin fluctuations in the T phase of $CaFe_2As_2$ are strongly suppressed or absent in the cT phase.

In Fig. 3 we compare the inelastic scattering data taken at (½ ½ 1), along both the [$H$ $H$ 0] direction and the [0 0 $L$] direction, after subtracting the empty can background, at points 2, 3 and 4. We also show the ambient pressure data taken at 180 K (point 1) in Fig. 1(c) where the background was estimated from the wings of the constant energy **q**-scan along [$H$ $H$ 0]. Using a single Gaussian line shape to fit the data in Fig. 3(a) we obtain a dynamic spin correlation length of 25 ± 5 Å at 3 meV along [$H$ $H$ 0], quite similar to the value obtained from similar measurements on $Ba(Fe_{0.92}Co_{0.08})_2As_2$.[2] The corresponding scan along [0 0 $L$] (Fig. 3(b)) is approximately 50% broader implying spin correlations perpendicular to the Fe-As planes with a shorter correlation length, again consistent with the findings of Lumsden *et al* [2]. Upon increasing the pressure to 0.5 GPa at 180 K (Fig. 3(c) and (d)), both constant energy **q**-scans evidence additional broadening, beyond that due to the small increase (0.4 deg) in the sample mosaic, indicating that the range of antiferromagnetic spin correlations decreases with increasing pressure. Figs. 3(e) and (f) show the background subtracted constant energy **q**-scans at 100 K in the cT phase (blue diamonds) and after heating the sample back to 180 K in the T phase (red diamonds). No clear structure is observed in the cT phase data along the [$H$ $H$ 0] direction (Fig. 3(e)), while a reduced and broadened peak at (½ ½ 1) is apparent in the re-measured data at 180 K. The increased width of the peak and reduced intensity is consistent with the increased sample mosaic and loss in sample volume. The corresponding re-measured scan along [0 0 $L$] at 180 K (Fig. 3(f)) is nearly transverse and, therefore, affected more by the increased sample mosaic. Nevertheless, scattering above the cT phase data can still be seen.

The observation of antiferromagnetically correlated spin fluctuations in the T phase, and their strong suppression or complete absence in the cT phase, provides interesting new perspectives concerning both the physical properties of $CaFe_2As_2$ and, more generally, superconductivity in the iron arsenides. Previous elastic neutron scattering studies [20, 21] have already provided strong evidence for the absence of static magnetic ordering in the cT phase. Total energy calculations [20] have shown that the Fe moment in the cT phase should be quenched. Additional electronic structure calculations using the local spin density approximation (LSDA) indicate that the cT phase is best characterized as a marginally itinerant system, on the brink of an instability for forming short- or long-range magnetic order. [24] Our results are consistent with the absence of static order and low-energy magnetic fluctuations in the cT phase at $(\frac{1}{2} \frac{1}{2} L)$ for $L$ odd. The possibility that these low-energy spin fluctuations are redistributed to other locations in momentum space can not be excluded, and deserves further exploration through both experiment and theory.

Transport measurements in $CaFe_2As_2$ under pressure show that there is a significant decrease in the resistivity as the sample transforms from the T phase to the cT phase. Although measurements performed with liquid pressure media [12, 15, 16] found a broad change in the resistivity, recent measurements using a He-pressure cell [19] observe sharp, discontinuous changes coincident with the first order T-cT structural transition determined by elastic neutron scattering experiments. [21] The differences in the measured transport properties appear to arise from the influence of a non-hydrostatic component in the liquid media/clamp cell measurements. The higher conductivity of the cT phase can, in the simplest model, be associated with either an increase in the electronic density of states, or a decrease in scattering (assuming that possible changes in Fermi velocities and phonon spectra are of lesser importance). Band structure calculations indicate a *decrease* in the electronic density of states (primarily Fe 3d contributions) at the Fermi energy when the sample transforms from the T to cT phase [21, 25], suggesting that electronic scattering effects play an important role in the increase in the cT phase conductivity. Here, we have shown that one source of scattering, arising from the presence of antiferromagnetically correlated spin fluctuations is reduced or eliminated in the cT phase.

The nature of the spin fluctuation contribution to electronic scattering certainly warrants further investigation, but it seems clear that the antiferromagnetic spin fluctuations in CaFe$_2$As$_2$ are coupled to the charge carriers, similar to what was found for the K-, Co-, and Ni-doped BaFe$_2$As$_2$ superconductors. Therefore, it is important to consider these results in the context of prior inelastic scattering measurements on the doped Ba$_{0.6}$K$_{0.4}$Fe$_2$As$_2$ [1], Ba(Fe$_{0.92}$Co$_{0.08}$)$_2$As$_2$ [2] and Ba(Fe$_{0.95}$Ni$_{0.05}$)$_2$As$_2$ [3] compounds. All of these studies have observed the opening of a gap in the spin fluctuation spectrum and the development of a resonant spin excitation at energies that scale with the superconducting transition temperature, $T_c$. The intensity of this resonance evolves as a power-law below $T_c$, a situation that is highly reminiscent of the high $T_c$ cuprate superconductors where it is widely held that spin fluctuations are strongly coupled to the superconducting order parameter. In light of recent transport measurements that show that, under hydrostatic pressure conditions, the cT phase itself does not support bulk superconductivity, the observed suppression or absence of both static antiferromagnetic order and dynamic spin fluctuations in the cT phase of CaFe$_2$As$_2$ lends support to the notion that spin fluctuations play a role for iron arsenide superconductivity as well.

We gratefully acknowledge the ILL for their rapid allocation of time and their support for this work. Work is supported by the U. S. Department of Energy Office of Science under the following contracts: at the Ames Laboratory under Contract No. DE-AC02 07CH11358 and at the Johns Hopkins Institute for Quantum Matter under Contract No. DE-FG02-08ER46544.

**Figure Captions**

**Fig. 1.** (color online) **(a)** Co-aligned single crystals of $CaFe_2As_2$. Approximately 300 crystals were mounted on both sides of 5 Al plates that were stacked to produce the measurement sample (shown on a millimeter grid); **(b)** A schematic of the reciprocal lattice plane probed in these measurements. The dashed lines denote scans along [$H\,H\,0$] and [$0\,0\,L$] directions; **(c)** The $p$-$T$ phase diagram of $CaFe_2As_2$ (after reference 21). The shaded area represents hysteretic regions and the numbers correspond to the measurement sequence as described in the text.

**Fig. 2.** (color online) Constant energy **q**-scans at 3 and 7 meV energy transfer through the (½ ½ 1) antiferromagnetic wave vector along the in-plane [$H\,H\,0$] direction at $T$ = 180 K and $p$ = 0.5 GPa for **(a)** and **(b)**; $T$ = 100K and $p$ = 0.5 GPa for **(c)** and **(d)**. Panels **(e)** and **(f)** show the data taken at $T$ = 180 K and $p$ = 0.5 GPa after passing through the T-cT transition twice (on cooling and warming). The open squares denote the empty can background measurements. The dashed vertical line denotes the position of the (½ ½ 1).

**Fig. 3.** (color online) Constant energy **q**-scans, after background subtraction, at 3 meV energy transfer through the (½ ½ 1) antiferromagnetic wave vector along the [$H\,H\,0$] (panels **(a)**, **(c)** and **(e)**) and [$0\,0\,L$] directions (panels **(b)**, **(d)** and **(f)**). The numbers in the shaded circles correspond to the points indicated in Fig. 1(c). The heavy lines through the data are fits to a single Gaussian line shape. The thin lines provide a guide to the eye. The dashed vertical line denotes the position of the (½ ½ 1). The instrumental resolution is shown at the bottom of panels **(a)** and **(b)**.

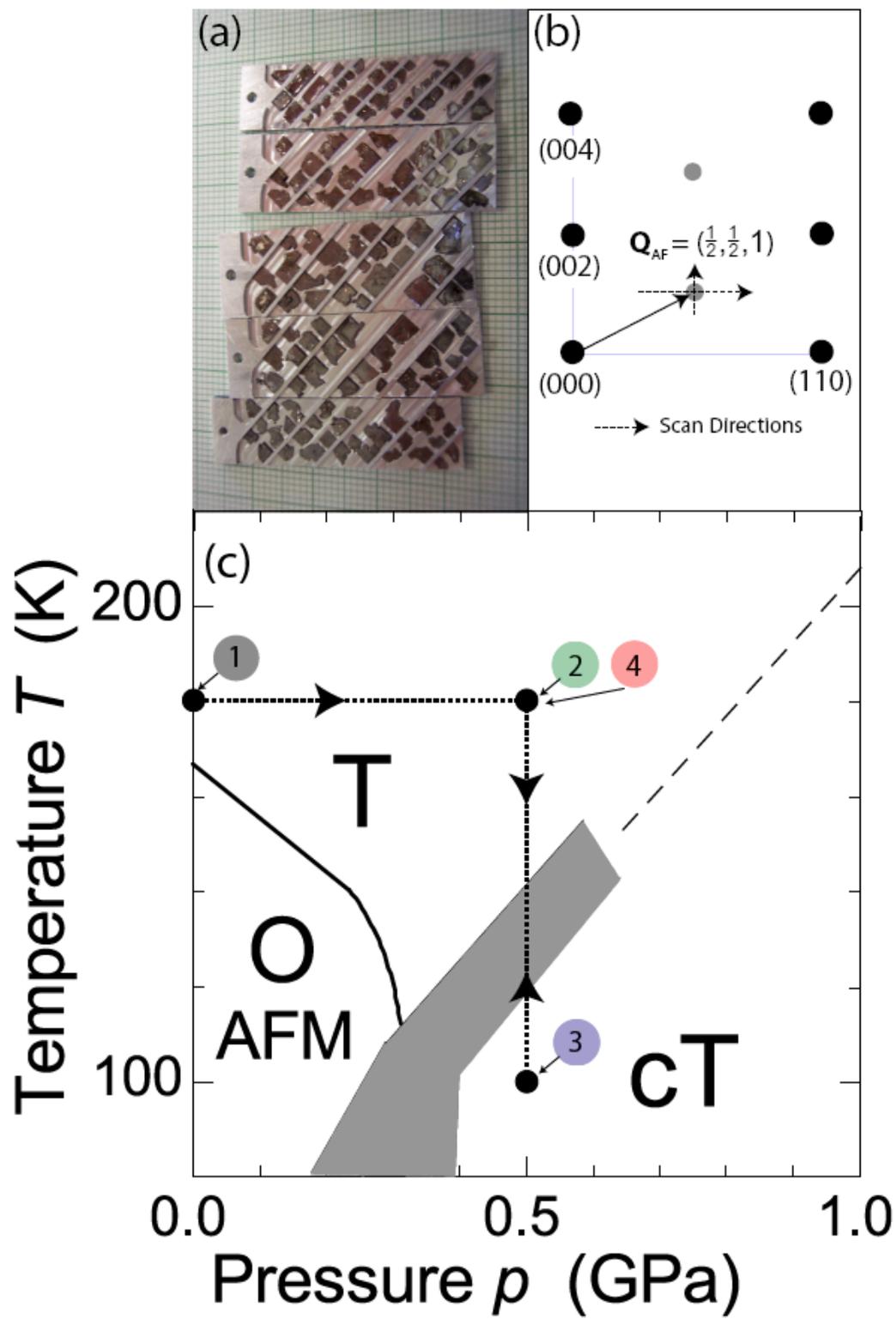

**Figure 1**

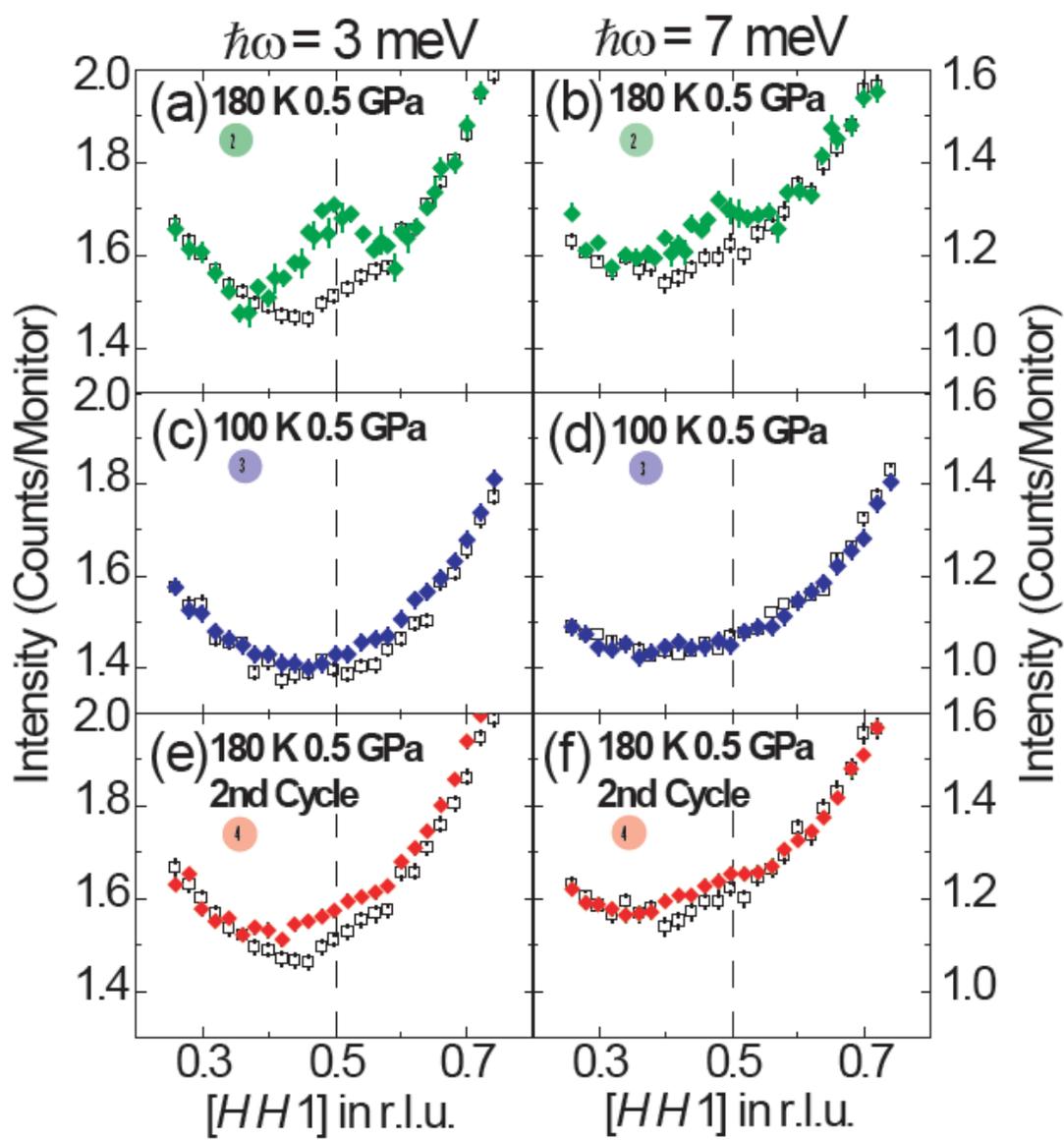

**Figure 2**

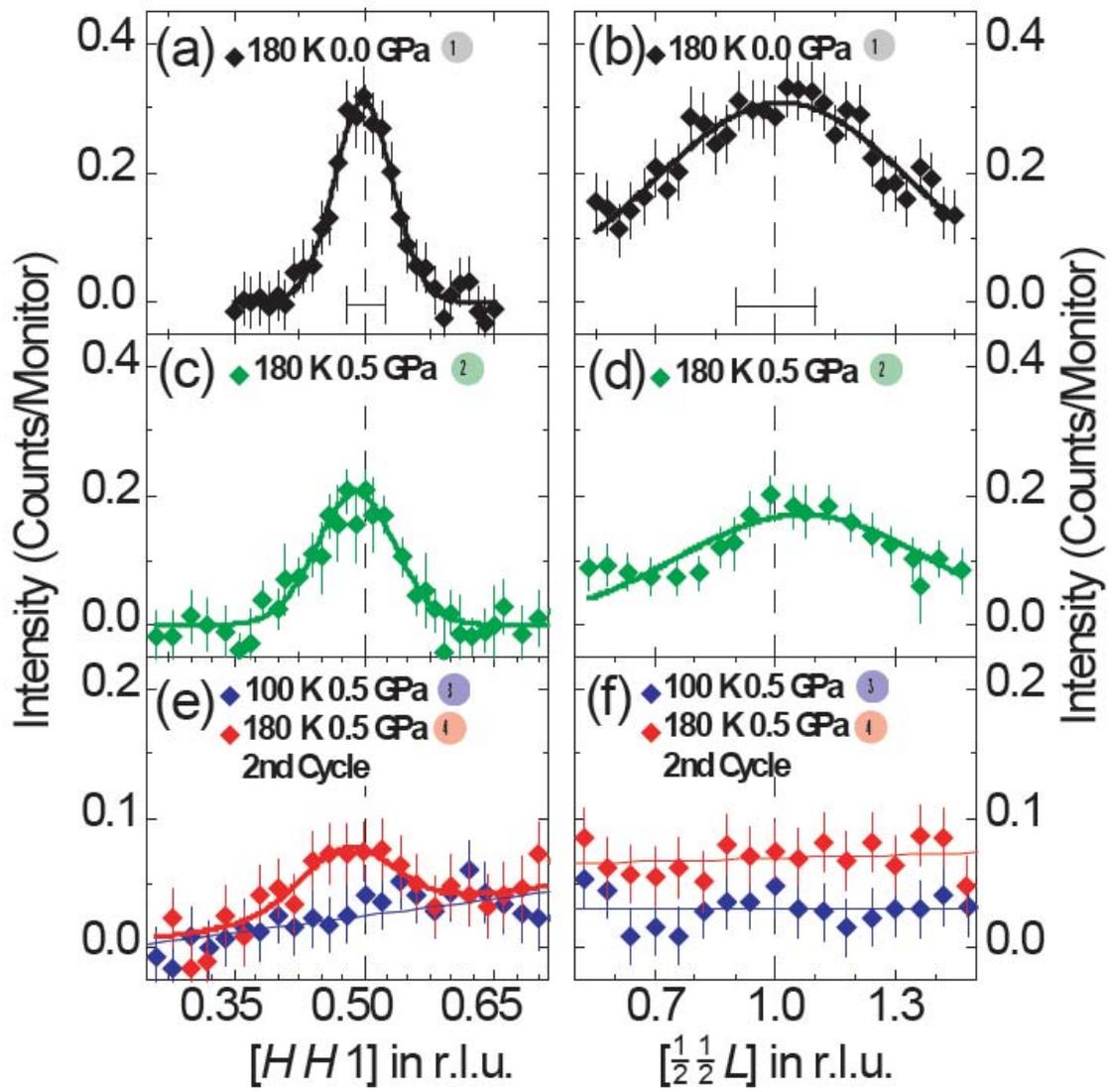

**Figure 3**